# Weak Ferroelectricity in *n* = 2 pseudo Ruddlesden-Popper-type niobate Li$_2$SrNb$_2$O$_7$

Takayuki Nagai,[1] Hirokazu Shirakuni,[2] Akitoshi Nakano,[2] Hiroshi Sawa,[3] Hiroki Moriwake,[4] Ichiro Terasaki,[2] and Hiroki Taniguchi[1,2]

[1]*Materials Research Center for Element Strategy, Tokyo Institute of Technology, Yokohama 226-8503, Japan*
[2]*Department of Physics, Nagoya University, Furocho, Chikusaku, Nagoya 464-8602, Japan*
[3]*Department of Applied Physics, Nagoya University, Furocho, Chikusaku, Nagoya 464-8603, Japan*
[4]*Nanostructures Research Laboratory, Japan Fine Ceramics Center, 2-4-1 matsuno, Atsutaku, Nagoya 456-8587, Japan*

**ABSTRACT:** Li$_2$SrNb$_2$O$_7$ (LSNO) crystallizes in a structure closely related to *n* = 2 Ruddlesden-Popper-type compounds, which is generally formed by intergrowth of 2-dimensional perovskite-type blocks and rocksalt-type layers. The present study demonstrates a coexistence of spontaneous polarization and anti-ferroelectric-like nonlinear response in LSNO at 80 K, suggesting a weak ferroelectricity below the phase transition temperature of 217 K. A combination of first-principles calculations and single crystal x-ray diffractions clarifies a polar *P*2$_1$*cn* structure for the ground state of LSNO, where an in-plane anti-ferroelectric displacement and an out-of-plane polar shift simultaneously take place. The present study offers a new perspective to design ferroelectric and antiferroelectric materials with Ruddlesden-Popper-type compounds.

## 1. INTRODUCTION

Ferroelectric materials are widely applied for electronic devices as non-volatile memories, infrared sensors, actuators, and so on [1, 2]. Designing of novel ferroelectric materials has therefore been an important issue both in fundamental science and materials engineering. The ferroelectric materials have conventionally been developed with perovskite-type oxides as represented by BaTiO$_3$ and PbTiO$_3$. Recently, *n* = 2 Ruddlesden-Popper-type (RP-type) oxides ($A_3B_2O_7$) have attracted increasing attention as new candidates for ferroelectric materials since a theoretical prediction of hybrid improper ferroelectricity (HIF) [3, 4]. In the HIF, spontaneous polarization can be induced by a combination of two nonpolar structural distortions, such as rotations and/or tilts of oxygen octahedra, without any support of a second-order Jahn-Teller effect [5-8], which causes polar distortions of coordination polyhedra. This concept has been experimentally verified in several compounds, for instance Sr$_3$Sn$_2$O$_7$ that possesses no second-order Jahn-Teller active cations [9, 10]. A hybrid improper antiferroelectricity, on the other hand, has also been proposed in Sr$_3$Zr$_2$O$_7$, suggesting a rich functionality in RP-type oxides [11]. Most recently we noticed a report of the antiferroelectricity in Li$_2$SrNb$_2$O$_7$, the same compound focused in the present study [12]. The antiferroelectricity in Li$_2$SrNb$_2$O$_7$, however, is incompatible with some of the experimental results, whereby careful reconsideration is neccesary.

Figures 1a and 1b show crystal structures of Ca$_3$Mn$_2$O$_7$ and Li$_2$SrNb$_2$O$_7$, respectively. Ca$_3$Mn$_2$O$_7$ crystallizes in the *n* = 2 RP-type structure, which consists of CaMnO$_3$ perovskite-type building blocks stacking along a [001] direction and CaO rocksalt-type layers inserted in every two perovskite-type layers [13, 14]. This material shows HIF, which is induced by two octahedral rotations of mutually different symmetry [3]. As shown in Fig. 1a, MnO$_6$ octahedra rotate and tilt around [001] and [110] axes, which are represented by $a^0a^0c^+$ and $a^-a^-c^0$ with the Grazer notation, respectively [15]. The combination of these octahedral rotations induces the ferroelectricity with a polar *A*2$_1$*am* structure in Ca$_3$Mn$_2$O$_7$. Li$_2$SrNb$_2$O$_7$, on the other hand, crystallizes in the structure closely related to the *n* = 2 RP-type structure [16, 17], where a configuration of the rocksalt layers is modified; Li$_2$O-layers lie instead of the rocksalt-type layers, resulting in oxygen deficiency at edges of the perovskite-type blocks. Note that LSNO has a centrosymmetric *Cmcm* structure with the oxygen octahedral tilting of $a^-a^-c^0$ at room temperature [16].

Here we show simultaneous appearance of ferroelectricity and anti-ferroelectricity along mutually perpendicular directions in LSNO, indicating weak ferroelectricity, which stems from close competition between antiferroelectric and ferroelectric instabilities [18-20]. The small spontaneous polarization of approximately 0.6 μC/cm$^2$ has been observed in *P-E* hysteresis measurements, being consistent with weak polar distortion in the *P*2$_1$*cn* structure that has been determined by a combination of single crystal x-ray diffractions and first-principle calculations. The present result sheds new light on fundamental understanding and materials designing of the layered-perovskite-type compounds.

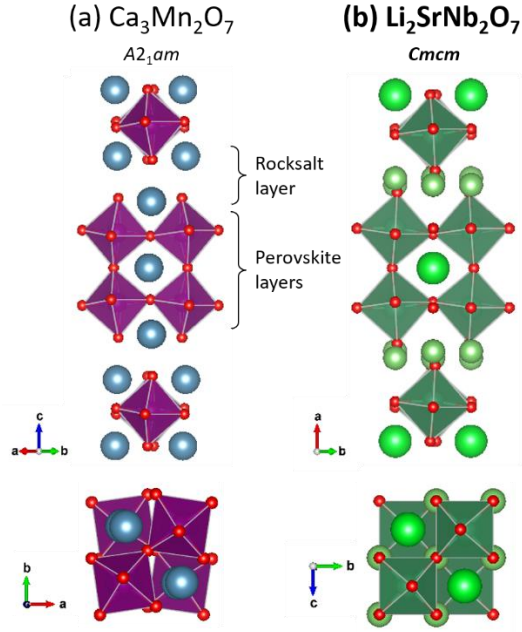

Figure 1. The crystal structure of (a) $Ca_3Mn_2O_7$ and (b) $Li_2SrNb_2O_7$ at room temperature, whose space groups at room temperature are $A2_1am$ and *Cmcm*, respectively. Blue spheres and Purple octahedra for $Ca_3Mn_2O_7$ correspond to Ca and $MnO_6$, respectively. Light green spheres, deep green spheres, and green octahedra for $Li_2SrNb_2O_7$, on the other hand, correspond to Li, Sr, and $NbO_6$, respectively.

## 2. EXPERIMENTAL SECTION

The polycrystalline samples of LSNO were prepared by a conventional solid-state reaction method with powder mixtures of $Li_2CO_3$ (99.9%), $SrCO_3$ (99.9%), $Nb_2O_5$ (99.99%), where excess $Li_2CO_3$ (10% mol) was added to compensate volatility of $Li_2O$ at high temperatures. The powder mixtures were ground in an agate mortar, and calcined in an alumina boat at 600°C for 12 h in air. After decarbonation, the products were reground, pelletized, and sintered at 1100°C for 6h in air. A single crystal (60 × 40 × 5 µm³) measured in the present study was picked from the polycrystalline pellet. The obtained powder and single crystal samples were respectively characterized by powder and single x-ray diffraction (XRD) measurements using BL02B2 and BL02B1 beamlines at SPring-8 with approvals of the Japan Synchrotron Radiation Research Institute (JASRI). The wavelengths of incident x-rays used in the present study were 16 keV (0.77 Å) for powder XRD and 50 keV (0.248 Å) for single crystal XRD measurements. According to the synchrotron powder XRD measurements, the obtained samples were found to include a tiny fraction of the impurity phase of approximately 5 %, which was assigned to paraelectric $Li_3NbO_4$. Dielectric permittivity and *P-E* hysteresis loop of the samples were measured using an 4284A precision LCR meter (Hewlett-Packard) and FCE-10 ferroelectric tester (Toyo Corporation), respectively. First-principles calculations were performed within the framework of density functional theory (DFT) using the projector-augmented wave (PAW) method [21] as implemented in the VASP code [22, 23]. A plane-wave cutoff energy of 500 eV and a 4 × 4 × 4 *k*-points mesh were used.

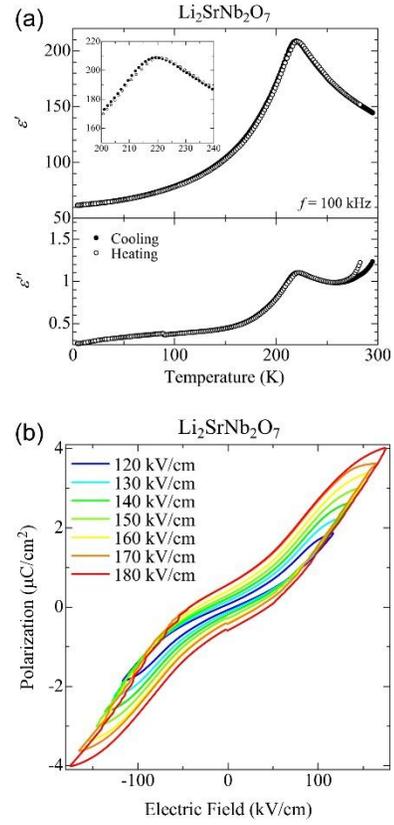

Figure 2. (a) Temperature dependence of the real part $\varepsilon'$ (top panel) and the imaginary part $\varepsilon''$ (bottom panel) of complex relative permittivity of polycrystalline LSNO, which were measured with a test frequency of 100 kHz. The full and open circles denote cooling and heating processes, respectively. Inset shows a magnified view for the temperature region around the phase transition temperature (b) *P-E* hysterisis loop of polycrystalline LSNO measured at 80 K with several muximum electric fields.

## 3. RESULTS AND DISCUSSION

Figure 2a shows the temperature dependence of the real part ($\varepsilon'$) and the imaginary part ($\varepsilon''$) of complex relative permittivity of polycrystalline LSNO, which were measured with a test frequency of 100 kHz. The real part $\varepsilon'$ increases gradually on cooling, and culminates at 217 K ($\equiv T_c$), indicating the phase transition. A maximum value of $\varepsilon'$ reaches around 200 at $T_c$, which is 3 times greater than that reported in Ref. [12]. This discrepancy would be due to difference in sample quality. The imaginary part $\varepsilon''$ also shows a small dielectric anomaly at the same temperature. Note that the value of $\tan\delta$ ($\equiv \varepsilon''/\varepsilon'$) is less than 1 % even at room temperature, indicating sufficient quality of the present sample to discuss the polarization response quantitatively. As shown in the inset of Fig. 2a, there is slight thermal hysteresis of approximately 2 K between the peak-like dielectric anomalies observed in cooling and heating processes. This result indicates that the phase transition at $T_c$ is of weak first-order.

To further investigate a nature of the phase transition at $T_c$, P-E hysteresis loop measurements have been performed at 80 K, which is sufficiently lower than $T_c$. As shown in Fig. 2b, the result exhibits a strongly non-linear behavior, which is composed of two distinct features: the first is an antiferroelectric-like double hysteresis loop, and the other is growth of remnant polarization as maximum applied field increases. The double hysteresis loop can be seen more clearly with the maximum applied field of 120 kV/cm due to the very small remnant polarization. The observed behavior is consistent with that reported in Ref. [12], the maximum applied field in which is slightly larger than 100 kV/cm. With increasing the applied field, on the other hand, the remnant polarization finally reaches to approximately 0.6 μC/cm². The maximum-applied-field-dependent remnant polarization is often observed in so-called minor loops, which appear when the maximum applied field is insufficient to saturate the spontaneous polarization. Nevertheless, the present result clearly demonstrates the polar symmetry of LSNO in the temperature region below $T_c$. Note that, though the system is not fully polarized, the spontaneous polarization of LSNO can be expected to be of the order of $10^0$ μC/cm². This value is much smaller than that of conventional proper ferroelectric oxides [24, 25]. Such a small spontaneous polarization is seen in improper ferroelectric oxides, as reported in $Gd_2(MoO_4)_2$ [26], Aluminate sodalites [27], and multiferroic $TbMnO_3$ [28]. The phase transition at $T_c$ is thus not the proper ferroelectric phase transition, whereby the spontaneous polarization is triggered by the other phase transition.

Figures 3a and 3b show diffraction patterns observed at temperatures lower and higher than $T_c$. In the high-temperature phase, the diffraction pattern can be reasonably indexed using the reported lattice constants with space group of *Cmcm* [15, 29-31]. At the low-temperature phase, on the other hand, new superlattice reflections violate the generation rule of *C*-centered lattice symmetry, *hkl*: $h + k = 2n$, for instance 20 5 5 and 22 5 5 in Fig. 3b. This result has demonstrated primitive orthorhombic symmetry of the low-temperature phase below $T_c$. It is noteworthy that, since intensities of superlattice reflections are of $10^{-2}$ of magnitude compared with fundamental reflections, relative atomic displacements associated with the phase transition are not so small. Judging from the generation rules we have found at low temperatures and the remnant polarization clarified by the *P-E* hysteresis measurements, the most plausible candidate for the crystal symmetry below $T_c$ is $P2_1cn$.

Here, we briefly discuss the structural change between high- and low-temperature phases according to results obtained by present structural analyses with the $P2_1cn$ symmetry, which are summarized in supplementary Tables S1 and S2. Figures 3e and 3f show subtracted $[SrNb_2O_7]^{2-}$-layers, which correspond to the components surrounded by dashed lines in Figs. 3c and 3d, respectively. In the high-temperature phase, there are mirror planes perpendicular to the *c*-axis (solid lines in Fig. 3e), whereas the mirror planes disappear in the low-temperature phase (see Fig. 3f). The disappearance of mirror symmetry is due to the relative displacements of Sr and Nb ions along the

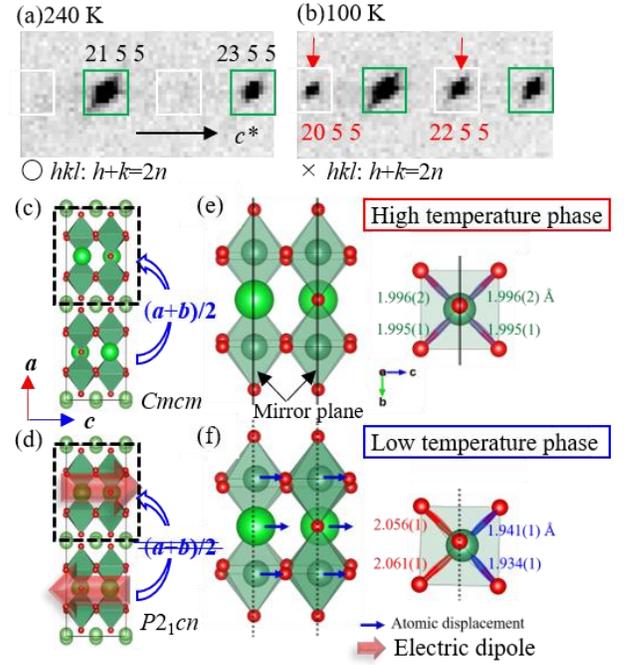

Figure 3. Diffraction patterns of LSNO observed at (a) 240 K and (b) 100 K. The obtained crystal structures of LSNO at 250 K and 30 K are respectively illustrated in (c) and (d), where the illustrations are viewed from the *b* axis. Arrows indicate the local dipole moments expected from the structural analysis. Local structures of $[SrNb_2O_7]^{2-}$, which are surrounded by dashed lines in (c) and (d), are shown in detail in (e) and (f), respectively.

[001] direction, which are indicated by blue arrows in Fig. 3f. The magnitude of the atomic displacements is ~ 0.1 Å, which is comparable with that observed in the ferroelectric phase transition of proper ferroelectric $BaTiO_3$. Such a relatively large atomic displacement along *c*-axis is consistent with relatively intense superlattice reflections. The lengths of Nb-O bonds in the $NbO_6$ octahedron differ by approximately 3% between the high-temperature and low-temperature phases, implying covalency between Nb and O plays a role in the phase transition at $T_c$. The relative displacements of Nb with respect to the oxygen octahedron can be regarded as a formation of electric dipole per one $[SrNb_2O_7]^{2-}$ layer along [001] direction as shown by red arrows in Fig. 3d. Since the *n*-glide plane perpendicular to the *c*-axis generates the electric dipole in the neighboring $[SrNb_2O_7]^{2-}$ layers, net polarization along [001] direction is zero. This configuration would give the antiferroelectric component observed in the *P-E* hysteresis measurements. Considering from the point group of $2mm$ for $P2_1cn$ symmetry, spontaneous polarization should be induced along [100] direction, which is perpendicular to the direction of electric dipoles induced by the relative displacements of Nb with respect to the oxygen octahedron. Note that the spontaneous polarization along [100] direction is expected to be quite small, because little difference was found in reliability factors for structural analyses with space groups of $P2_1cn$ and its centrosymmetric super group, *Pmcn*.

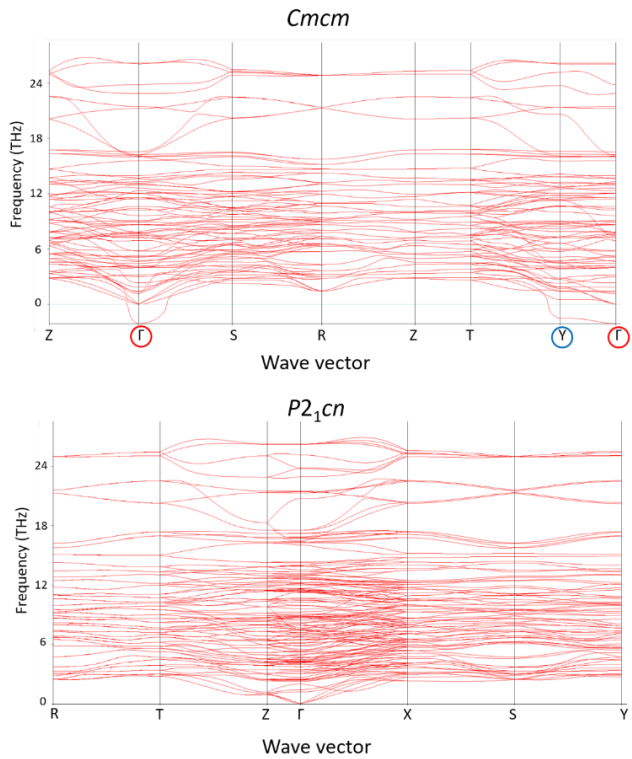

Figure 4. Phonon dispersion curves of LSNO calculated with *Cmcm* (top) and *P*2$_1$*cn* (bottom) phases. Circles indicate the softmodes at the Γ- and Y-points of Brillouin Zone.

To elucidate the mechanism of phase transition in LSNO, we have investigated phonon dynamics of LSNO by first principles calculations. The upper panel of Fig. 4 shows phonon dispersion curves for the high temperature centrosymmetric *Cmcm* phase at room temperature [28-30]. As shown in the figure, the calculation has clarified two softmodes at Γ- and Y-points, which have almost same imaginary frequencies. Within a framework of second-order phase transition, freezing of the Γ-point softmode gives polar *Cmc*2$_1$ phase, leading to the proper ferroelectric phase transition. The Y-point softmode, on the other hand, induces the phase transition into non-polar *Pmcn* structure, which satisfies a necessary condition for the antiferroelectric phase transition. The calculated phonon dispersion thus indicates the close competition between ferroelectric and anti-ferroelectric instabilities in LSNO similarly to the case of Bi$_2$SiO$_5$ [32, 32]. Since the present x-ray diffraction measurements have clarified the appearance of super structure reflections at $T_c$, the *C*-centered lattice can be automatically ruled out from candidates for the crystal symmetry below $T_c$ as discussed in the preceding section. Thus, the phonon instability, which drives the phase transition at $T_c$, is the Y-point softmode, though the imaginary frequency of Γ-point softmode is slightly larger than that of the Y-point softmode, probably due to accuracy of the calculation. Note that freezing of the Y-point softmode never gives the polar low-temperature phase in the second-order phase transition. However, the dielectric measurements have suggested the weak first order-phase transition at $T_c$ as presented in the inset of Fig. 2a. By considering the present results comprehensively, the most plausible scenario for the phase transition at $T_c$ would be the weak ferroelectric phase transition as in the case of PbZrO$_3$; The ferroelectricity appears as a result of the close competition between the ferroelectric and antiferroelectric instabilities, whereas the phase transition is intrinsically driven by the antiferroelectric instability [18]. Finally, Fig. 4b presents the phonon dispersion curves calculated with the polar *P*2$_1$*cn* symmetry. Since no softmode is obtained in the calculation as shown in the figure, the polar ground state of *P*2$_1$*cn* symmetry is reasonable for LSNO.

## 4. CONCLUSION

We have demonstrated the weak first-order phase transition of LSNO between the centrosymmetric *Cmcm* to the polar *P*2$_1$*cn* phases at $T_c$. In the low-temperature polar phase, the coexistence of ferroelectricity and antiferroelectricity has been found by the *P-E* hysteresis measurements. The structural analyses using the single crystal x-ray diffractions have clarified the mutually perpendicular configuration of ferroelectric and antiferroelectric polarization directions. The first-principles calculations have confirmed the ground state of polar *P*2$_1$*cn* symmetry in LSNO, and have elucidated the competing lattice instability at Γ- and Y-points. The present results indicate the weak ferroelectricity of LSNO, where the spontaneous polarization is triggered by the antiferroelectric phase transition through the competing ferroelectric and antiferroelectric lattice instabilities. The present finding updates fundamental understanding of the phase transition in layered-perovskite-type compounds, and provides a new insight into designing functional ferro/antiferroelectric materials.

## ACKNOWLEDGMENT

The present work was partially supported by JST CREST Grant Number JPMJCR17J2, and the approvals of the Japan Synchrotron Radiation Research Institute (JASRI) (Proposals No. 2018B1713 and No. 2018A1653). One of the authors (T. Nagai) was supported by the Program for Leading Graduate Schools "Integrative Graduate Education and Research in Green Natural Sciences", MEXT, Japan and a Grant-in-Aid for JPSJ Research Fellow (No. 18J11905), MEXT, Japan.